\documentclass[prb,onecolumn]{revtex4-1}

\usepackage{graphicx}
\normalfont
\usepackage{amsmath}
\usepackage{upgreek}
\usepackage{epstopdf}

\usepackage{epstopdf}

\newcommand{\micro}{${\upmu}$}

\begin{document}

\title{Exciton-polaritons in van der Waals heterostructures embedded in tunable microcavities}

\author{S. Dufferwiel$^{1\ast}$}
\author{S. Schwarz$^{1\ast}$}
\author{F. Withers$^2$}
\author{A. A. P. Trichet$^3$}
\author{F. Li$^1$}
\author{M. Sich$^1$}
\author{O. Del Pozo-Zamudio$^1$}
\author{C. Clark$^4$}
\author{A. Nalitov$^5$}
\author{D.D. Solnyshkov$^5$}
\author{G. Malpuech$^5$}
\author{K. S. Novoselov$^2$}
\author{J. M. Smith$^3$}
\author{M. S. Skolnick$^1$}
\author{D. N. Krizhanovskii$^1$}
\author{A. I. Tartakovskii$^1$}

\affiliation{$^1$Department of Physics and Astronomy, University of Sheffield, Sheffield S3 7RH, UK}

\affiliation{$^2$School of Physics and Astronomy, University of Manchester, Manchester M13 9PL, UK}

\affiliation{$^3$Department of Materials, University of Oxford, Parks Road, Oxford OX1 3PH, UK}

\affiliation{$^4$Helia Photonics, Livingston EH54 7EJ, UK}

\affiliation{$^5$ Institut Pascal, Blaise Pascal University,
24 avenue des Landais, 63177 Aubi\'ere Cedex, France}

\author{$\ast$ These authors contributed equally to this work.}

\date{\today}

\begin{abstract}
Layered materials can be assembled vertically to fabricate a new class of van der Waals (VDW) heterostructures a few atomic layers thick, compatible with a wide range of substrates and optoelectronic device geometries, enabling new strategies for control of light-matter coupling. Here, we incorporate molybdenum diselenide/boron nitride (MoSe$_2$/hBN) quantum wells (QWs) in a tunable optical microcavity. Part-light-part-matter polariton eigenstates are observed as a result of the strong coupling between MoSe$_2$ excitons and cavity photons, evidenced from a clear anticrossing between the neutral exciton and the cavity modes with a splitting of 20 meV for a single MoSe$_2$ monolayer QW, enhanced to 29 meV in MoSe$_2$/hBN/MoSe$_2$ double-QWs. The splitting at resonance provides an estimate of the exciton radiative lifetime of 0.4 ps. Our results pave the way for room temperature polaritonic devices based on multiple-QW VDW heterostructures, where polariton condensation and electrical polariton injection through the incorporation of graphene contacts may be realised.
\end{abstract}

\maketitle

Recently, the fabrication of vertical assemblies of two-dimensional (2D) materials has become possible providing novel types of heterostructures with controlled and tunable properties \cite{Dean2010,Britnell2013,Geim2013}. The weak interlayer bonding allows a variety of 2D layers with different lattice constants to be stacked on top of one another, creating artificial materials with new material characteristics \cite{Geim2013}. The potential applications of these VDW heterostructures are further widened through the incorporation of semiconducting transition metal dichalcogenide (TMDC) monolayers  \cite{MakPRL2010,NovoselovPNAS2005,WangNatNano2012,XuNatPhys2014}. Unlike their bulk form, monolayer TMDCs are direct band-gap semiconductors \cite{MakPRL2010} exhibiting pronounced exciton resonances at room temperature owing to the exceptionally high exciton binding energies of a few 100 meV \cite{Ugeda2014,Berkelbach2013} as well as displaying coupled spin and valley degrees of freedom\cite{Mak2012}. Recently, electroluminescence from lateral p-n junctions has been demonstrated based upon hBN-WSe$_2$ heterostructures\cite{Ross:2014fk} and the stacking of TMDC layers has been shown to lead to the formation of long-lived interlayer excitons \cite{Rivera2015}. In more complex heterostructures incorporating hBN barriers and graphene electrodes, efficient electroluminescence was observed from WS$_2$, MoS$_2$ and WSe$_2$ monolayers under vertical current injection \cite{Withers2015}. 

In this work, we demonstrate how the incorporation of VDW heterostructures into optical microcavities enables control and modification of the light-matter interaction. We place MoSe$_2$/hBN heterostructures in open tunable cavities having high reflectivity dielectric mirrors with adjustable separation \cite{Dufferwiel2014,Schwarz2014}. In contrast to previous experiments, where modification of the emission pattern and radiative recombination rate of 2D films coupled to cavities was observed \cite{Gan2013,Wu2014,Schwarz2014}, as well as a recent report on lasing in photonic crystal nanocavities \cite{Wu2015}, we present conclusive evidence for the strong light-matter interaction regime and the formation of part-light, part-matter polariton eigenstates. This regime is observed when the cyclic emission and reabsorption of light inside a microcavity occurs on a timescale faster than the exciton and photon dissipation rates, a regime favoured by the large oscillator strength of the direct-band-gap optical transition in TMDC monolayers \cite{Cheiwchanchamnangij2012,Ugeda2014,Jiang2014}. Here we evidence strong coupling in reflectivity and photoluminescence (PL) through the anticrossing of the tunable cavity mode energy and the MoSe$_2$ exciton energy showing the formation of upper and lower polariton branches. A large Rabi splitting of 20 meV is observed for a heterostructure containing a single MoSe$_2$ monolayer, whereas this splitting is increased to 29 meV for a multiple quantum well (MQW) structure with two MoSe$_2$ monolayers separated by a hBN layer. From the coupling strength we extract a radiative exciton lifetime of 0.4 ps corresponding to a homogeneous linewidth of 1.6 meV. In contrast to previous work in TMDC microcavities, where the observed spectral features are poorly resolved preventing an unambiguous claim of strong coupling,\cite{Liu2015} we show fully resolved polariton branches with a Rabi splitting that significantly exceeds the polariton linewidths. We also observe a notable difference between the exciton-photon coupling for the neutral (X$^0$) and negatively charged (X$^-$) excitons, with an intermediate coupling regime observed for $X^-$, consistent with its reduced oscillator strength, providing further insight into the nature of the electronic states in TMDC monolayers. Moreover, the large exciton binding energy of TMDCs allows the observation of a narrow upper polariton branch with comparable intensity to the lower polariton at resonance. This arises due to the significant separation of the electron-hole continuum from the upper polariton states, potentially opening a new regime of polariton physics.

The study of exciton-polaritons has revealed a wealth of rich phenomena such as Bose-Einstein condensation in the solid state \cite{Kasprzak2006}, polariton superfluidity \cite{Amo2009_2}, as well as room temperature polariton lasing in the UV and blue spectral regions using wide band gap materials such as GaN \cite{Christopoulos2007,Bhattacharya2014} and ZnO \cite{Lu2012,Li2013} or organic materials \cite{KenaCohen2010,Daskalakis2014,Plumhof2014}. The integration of TMDC heterostructures in optical microcavities is an attractive alternative to previously studied systems. TMDCs exhibit very large exciton binding energies and sharp exciton resonances, whose properties can be tailored by combining a wide variety of 2D crystals in heterostructures. Observation of room temperature excitons in TMDCs combined with the recently demonstrated good electroluminescence of VDW heterostructures \cite{Withers2015} lays the foundation for the development of low threshold electrically pumped polariton lasers operating in the visible and near infrared. These devices can be easily incorporated onto a wide range of substrates allowing the development of hybrid TMDC/III-V microcavity structures as well as electrically driven polariton devices with vertical current injection using graphene contacts. This work opens the way to these developments in the rich material system of VDW crystals.

\begin{figure*}[htdp]
\includegraphics[width=\textwidth]{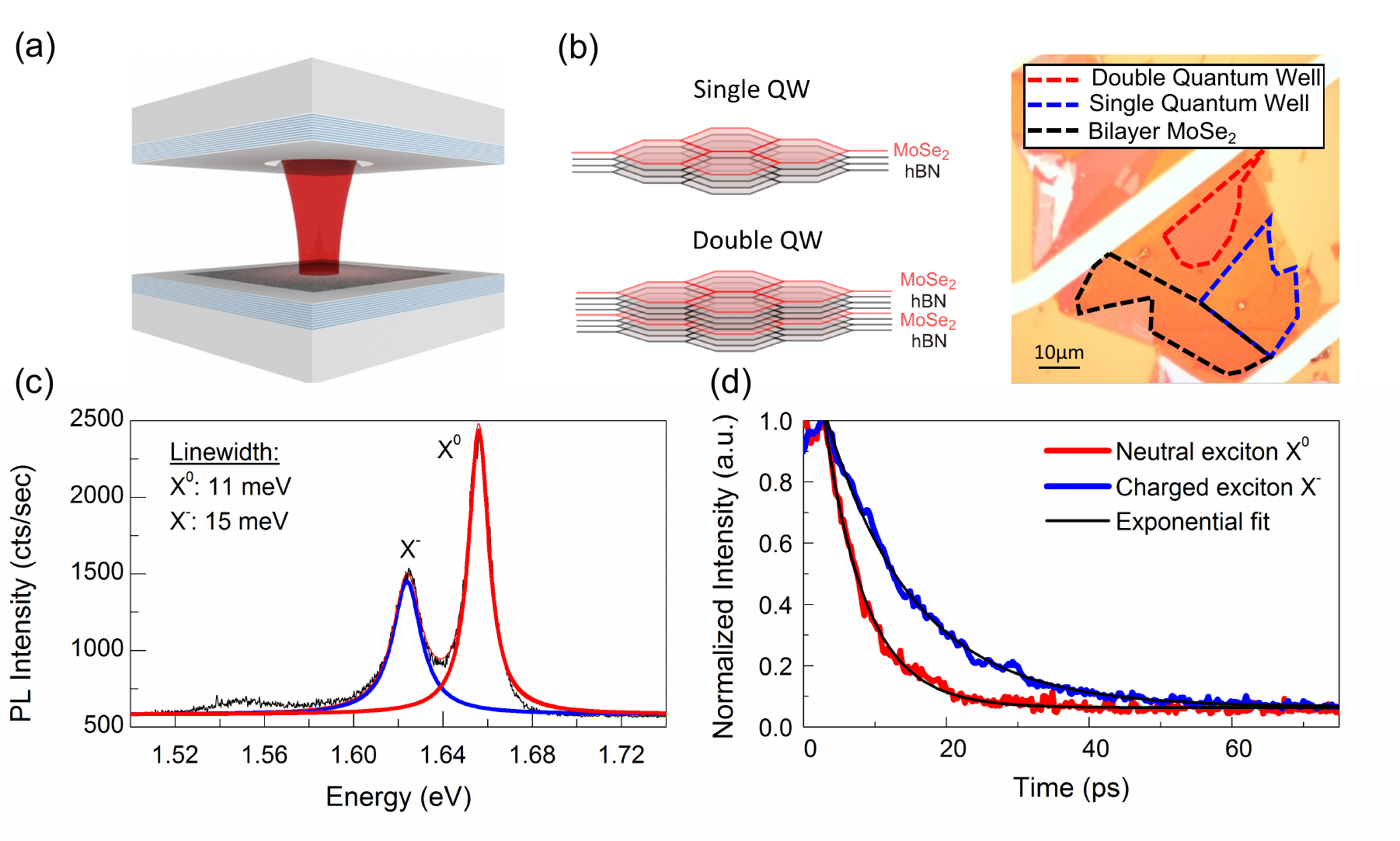}
\caption{\label{fig1} {\bf MoSe$_2$ heterostructures embedded in a tunable open-access microcavity.} (a) Schematic of the tunable hemispherical cavity with an embedded MoSe$_2$ heterostructure. (b) The left panel shows a schematic of the single and double QW heterostructures. The right panel displays an optical image of the MoSe$_2$ heterostructure where the single and double QW areas are marked by the blue and red boxes respectively.  A bilayer MoSe$_2$ region is marked by the black lines. The structure is fabricated on the surface of the planar DBR at an electric-field antinode of the formed microcavity. (c) The PL emission of a monolayer of MoSe$_2$ at 4 K shows two characteristic peaks attributed to a neutral (X$^0$) and charged (X$^-$) exciton with a measured linewidth of 11 meV for X$^0$ and 15 meV for X$^-$. (d) Time-resolved measurements reveal a PL lifetime of 5.3 ps for X$^0$ and 12.5 ps for X$^-$.}
\end{figure*}

The tunable microcavity is formed by one planar and one concave dielectric distributed Bragg reflector (DBR)\cite{Dolan2010}. A schematic of the formed microcavity is shown in Figure~\ref{fig1}(a) where two nanopositioner stacks allow the independent positioning of the two DBRs and the cavity mode resonances can be tuned in-situ through control of the mirror separation. The heterostructure is fabricated through standard mechanical exfoliation and then transferred to the surface of the planar DBR, at an electric field antinode of the formed microcavity. The heterostructure consists of different areas corresponding to 3 different active regions; a single QW region, a double QW region and a bilayer region. The single QW region consists of a single monolayer sheet of MoSe$_2$ placed on a $3$ nm thick sheet of hexagonal boron nitride (hBN) as outlined by the blue dashed lines in the optical image in Figure~\ref{fig1}(b). The double QW structure consists of two monolayer sheets of MoSe$_2$, separated by a $3$ nm film of hBN, placed on a thin sheet of hBN outlined by the red dashed lines in Figure~\ref{fig1}(b). Additionally an area of a bilayer MoSe$_2$ is enclosed within black dashed lines. The low temperature PL spectrum from a single monolayer of MoSe$_2$ is shown in Figure~\ref{fig1}(c). The spectrum consists of a neutral exciton X$^0$ and a negatively charged trion X$^-$ with linewidths of 11 meV and 15 meV respectively.\cite{RossNatComm2013} Time-resolved measurements of the PL are shown in Figure~\ref{fig1}(d) revealing a lifetime of 5.3 ps for $X^0$ and 12.5 ps for $X^-$, consistent with those reported elsewhere.\cite{Wang2012}

\begin{figure*}
\includegraphics [width=\textwidth] {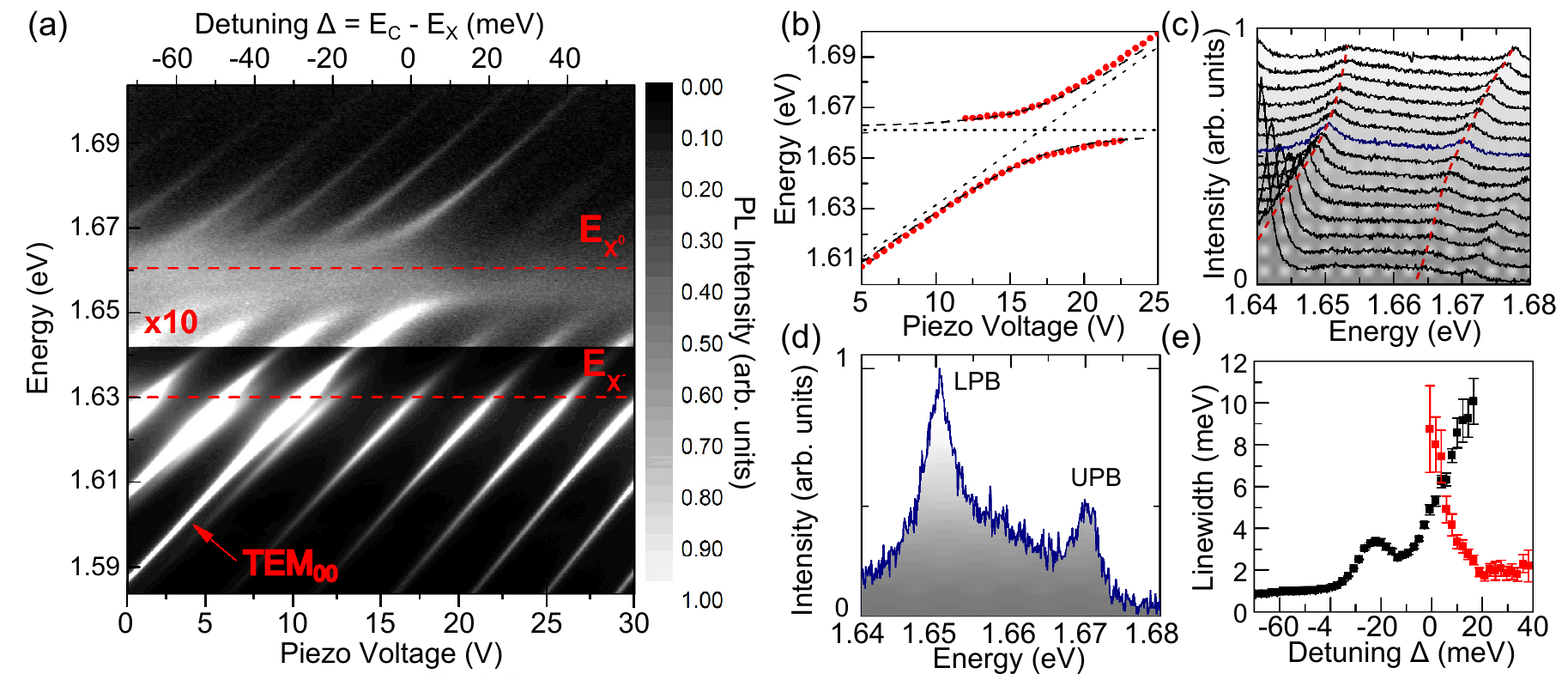}
\caption{\label{fig2} {\bf Observation of strong exciton-photon coupling in a MoSe$_2$ single QW heterostructure.} (a) A clear anticrossing in PL is observed between the discrete cavity mode energies and the neutral exciton energy at 4 K. (b) The upper and lower polariton branches are fitted for the longitudinal mode with a vacuum Rabi splitting of 20 meV. The longitudinal resonance is labelled TEM$_{00}$. (c) PL spectra of the longitudinal mode at various exciton-photon detunings from $\Delta = -16$ meV (bottom) to $\Delta = +12$ meV (top). (d) The PL spectrum on resonance shows the UPB and LPB well resolved. (e) The linewidth of the LPB and UPB as a function of detuning. On resonance the linewidth for LPB and UPB is around 4.8 meV and 8.5 meV. At a negative detuning of $-25$ meV the LPB is broadened due to the intermediate coupling with the charged exciton state.}
\end{figure*}

\begin{figure}
\includegraphics [scale=1.0] {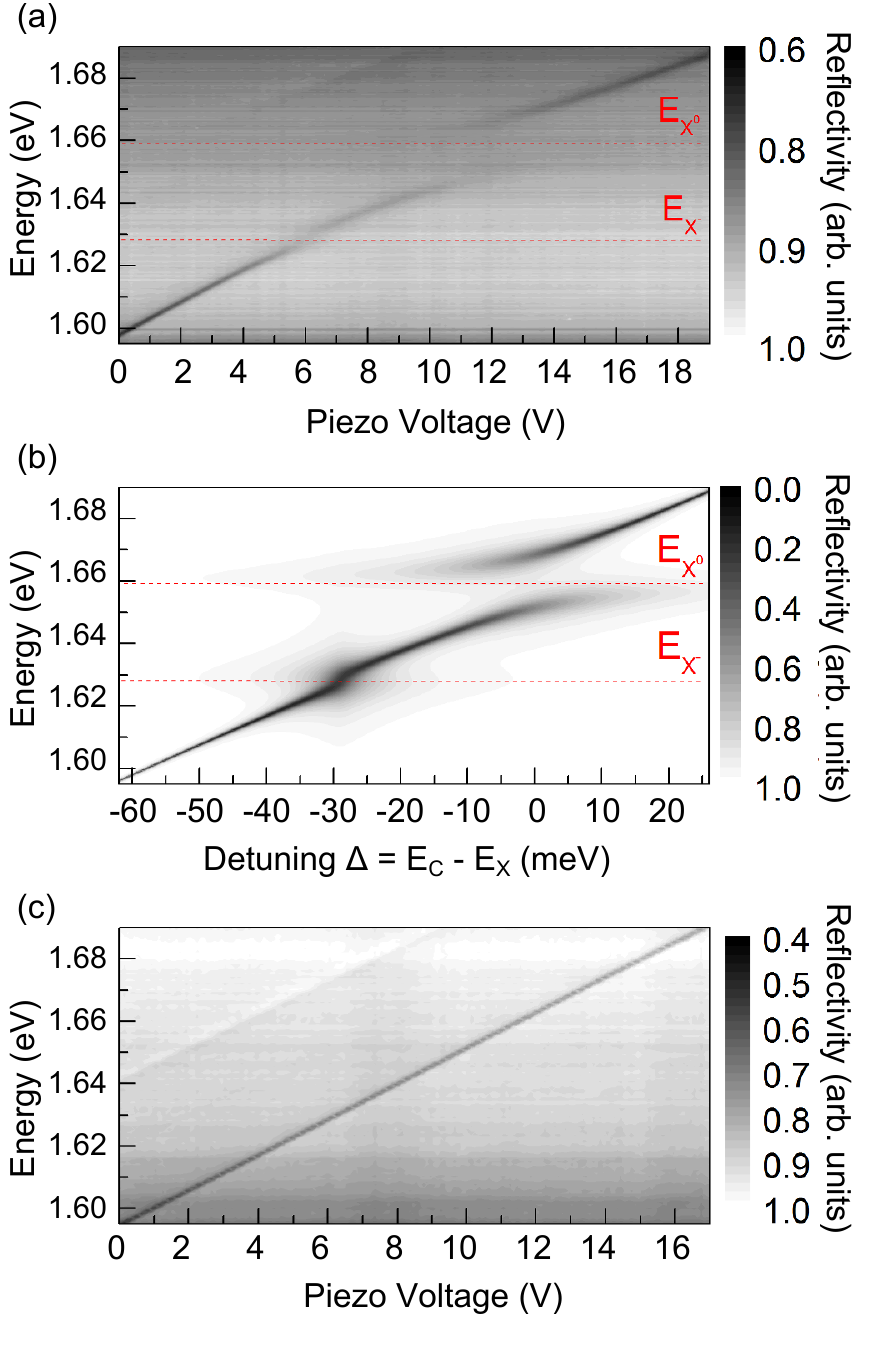}
\caption{\label{fig3} {\bf Intermediate exciton-photon coupling regime for $X^-$ observed in reflectivity measurements} (a) Reflectivity scan of single QW area at 4 K showing clear anticrossing with $X^0$. Intermediate coupling with the $X^-$ is also apparent when close to the $X^-$ resonance. (b) Theoretical reproduction of (a) based on a 3 coupled oscillator model with coupling strengths of 18 meV and 8.2 meV for $X^0$ and $X^-$ respectively. (c) Reflectivity scan of an empty cavity with no active region showing a linear dependence of the cavity mode energy on the piezo voltage.}
\end{figure}

The open cavity system allows spatial xyz-positioning of the two mirrors independently. As such, any area of the MoSe$_2$ heterostructure on the planar DBR can be placed in the optical path and a microcavity formed with the selected area as the active region. The micron sized Gaussian beam waist on the planar mirror of the formed cavity allows each region of the MoSe$_2$ heterostructure to be coupled to the cavity modes independently\cite{Dufferwiel2014}. Figure~\ref{fig2}(a) shows a PL scan of a cavity formed with a concave mirror with a radius of curvature of $20$ \micro m and a single monolayer MoSe$_2$ active region. The various modes present in the spectra arise from the three-dimensional confinement of the photonic field which gives rise to longitudinal modes and their associated higher order transverse modes. These cavity modes are tuned through the neutral exciton resonance by reducing the mirror separation by applying a DC voltage to the bottom z-nanopositioner. The longitudinal resonance, labelled TEM$_{00}$, is at an energy of 1.588 eV at $V = 0$ and the modes at higher energy are its associated first (1.608 eV) and second (1.628 eV) transverse modes. The total optical cavity length is around $2.3$ \micro m and the longitudinal mode number $q = 5$. The modes at lower energy than the longitudinal mode are transverse modes associated with a different longitudinal mode at much lower energy ($q-1$). These are present since the mirror separation is larger than the separation of $\lambda/2$ required to reach the fundamental longitudinal resonance ($q=1$). Clear anticrossings between the cavity mode resonances and the neutral exciton energy are observed revealing the formation of well-resolved polariton states. Each photonic mode is characterized by a specific field distribution in the plane of the TMDC layers and couples to an excitonic mode with the same in-plane distribution. As a result, different photonic modes couple to spatially orthogonal exciton states. Polariton states from different photon modes are therefore orthogonal and are well described by the coupling between a single photon mode and a single excitonic mode for each of them. A fit to the upper and lower polariton branches (UPB and LPB) peak energies for the longitudinal mode using a coupled oscillator model is shown by the dashed lines in Figure~\ref{fig2}(b) and reveal a Rabi splitting of $\hbar\Omega_{Rabi} = 20$ meV for a single MoSe$_2$ monolayer. Here we approximate the cavity mode energy as a linear function of applied voltage which is supported by both transfer matrix simulations and reflectivity measurements (see Figure~\ref{fig3}(c)). The detuning is then defined as $\Delta = E_{ph} - E_{X^0}$ where $E_{ph}$ and $E_{X^0}$ are the fitted longitudinal cavity mode energy (TEM$_{00}$) and the neutral exciton energy respectively. Spectral slices of the PL at various detunings from $\Delta = -16$ meV to $\Delta = +12$ meV are displayed in Figure~\ref{fig2}(c) where the LPB and UPB can be resolved. Moreover, the PL spectrum at zero detuning shown in Figure~\ref{fig2}(d) shows clearly two peaks where $\hbar\Omega_{Rabi}$ is significantly larger than the polariton linewidths $\gamma_{LPB}$ and $\gamma_{UPB}$. 

The polariton linewidths are plotted as a function of longitudinal mode detuning in Figure~\ref{fig2}(e). At large negative detunings of $\Delta < - 30$ meV, the LPB linewidth approaches the bare cavity linewidth of $0.8$ meV due to the high photonic component of the polariton. At detunings of $\Delta = - 30$ to $- 20$ meV significant broadening of the LPB is observed, corresponding to resonance with the X$^-$ energy. This broadening is attributed to intermediate coupling between the $X^-$ states and the cavity mode where the Rabi splitting is comparable to the corresponding polariton linewidths in PL but can be resolved only in reflectivity as discussed below. At zero detuning the LPB linewidth is $\gamma_{LPB} = 4.9 $ meV which is less than the linewidth averaged value of $(\gamma_{X^0}+\gamma_{ph})/2 = 5.9$ meV predicted from the two-level coupled oscillator model, where $\gamma_{X^0}$ and $\gamma_{ph}$ are the measured inhomogeneously broadened exciton linewidth and bare cavity linewidth taken from PL measurements. This may be due to motional narrowing which is expected in systems such as this where $\hbar \Omega_{Rabi} >> \gamma_{X^0}$ leading to averaging over the inhomogeneous broadening.\cite{Whittaker1998} This causes the polariton linewidth to approach $(\Gamma_{X^0}+\gamma_{ph})/2$ where $\Gamma_{X^0}$ is the homogeneous neutral exciton linewidth.\cite{Kavokin1998} 
Alternatively, this narrowing may be due to the reduced number of excitonic states that couple to the photonic mode within the 1 \micro m beam waist which may have a smaller linewidth in comparison to the 2 \micro m spot measured in Figure \ref{fig1}(c) due to disorder in the film. In contrast, the UPB linewidth at resonance is $8.7$ meV due to broadening from relaxation through scattering to the uncoupled exciton states\cite{Savona1997}. The radiative lifetime of the neutral exciton can be estimated from the Rabi splitting to be around $\tau = 0.4$ ps, corresponding to a homogeneous linewidth of $\Gamma_0 = 1.6$ meV in agreement with recent work in WSe$_2$ monolayers \cite{Moody2014} (see Supplementary Information).

The UPB is observed at positive detunings up to the recorded detunings of $\Delta = +40$ meV with a relatively narrow linewidth of around $2$ meV. This is possible in TMDCs since the binding energy, $E_B >> \hbar\Omega_{Rabi}$ and hence the electron-hole continuum is far from the polariton resonances leading to much reduced relaxation of the UPB states. This is a unique property of TMDCs which also allows both the LPB and UPB to have comparable intensities at resonance. This can be quantified further through the ratio $\hbar\Omega_{Rabi}/E_{B}$, which is around $\approx 0.04$ for MoSe$_2$ in contrast to $> 0.2$ for all other materials where strong coupling was observed. This allows the UPB resonance to be both bright and narrow while remaining on the mirror stopband  (see Supplementary Information).

Figure~\ref{fig3}(a) shows a reflectivity scan of a single QW area. Only the longitudinal mode is visible due to the poor mode matching between the Gaussian excitation spot and the lateral transverse mode profiles. When the cavity mode is tuned through resonance with the $X^-$ energy a small shift in the cavity energy can be observed, in contrast to the broadening observed in PL. This difference in behaviour can be understood through the dependence of the observed Rabi splitting on the measurement method, such as absorption, reflectivity or transmission, when the polariton linewidths are comparable to the mode splitting. In this case the splitting in reflectivity is expected to be larger than in PL ($\Omega_R > \Omega_{PL}$)\cite{Savona1995}. The $X^-$ coupling strength is proportional to $\sqrt{n_e}$ where $n_e$ is the electron density which is due to inherent doping from impurities in the exfoliated sample\cite{Rapaport2001}. In Figure~\ref{fig3}(b) we theoretically reproduce the experimental result shown in Figure~\ref{fig3}(a) based on a 3 coupled oscillator model. We use the exciton linewidths from Figure~\ref{fig1}(c) and reproduce the coupling behaviour using coupling strengths for $X^0$ and $X^-$ of $18$ meV and $8.2$ meV respectively. This is consistent with the reduced oscillator strength of $X^-$ evidenced through its low absorption.\cite{RossNatComm2013} It has been shown in GaAs based systems that the total oscillator strength of all excitonic components is a conserved quantity and hence the presence of X$^-$ due to doping reduces the coupling strength of X$^0$.\cite{Rapaport2001} 

\begin{figure}
\includegraphics[scale=1.0]{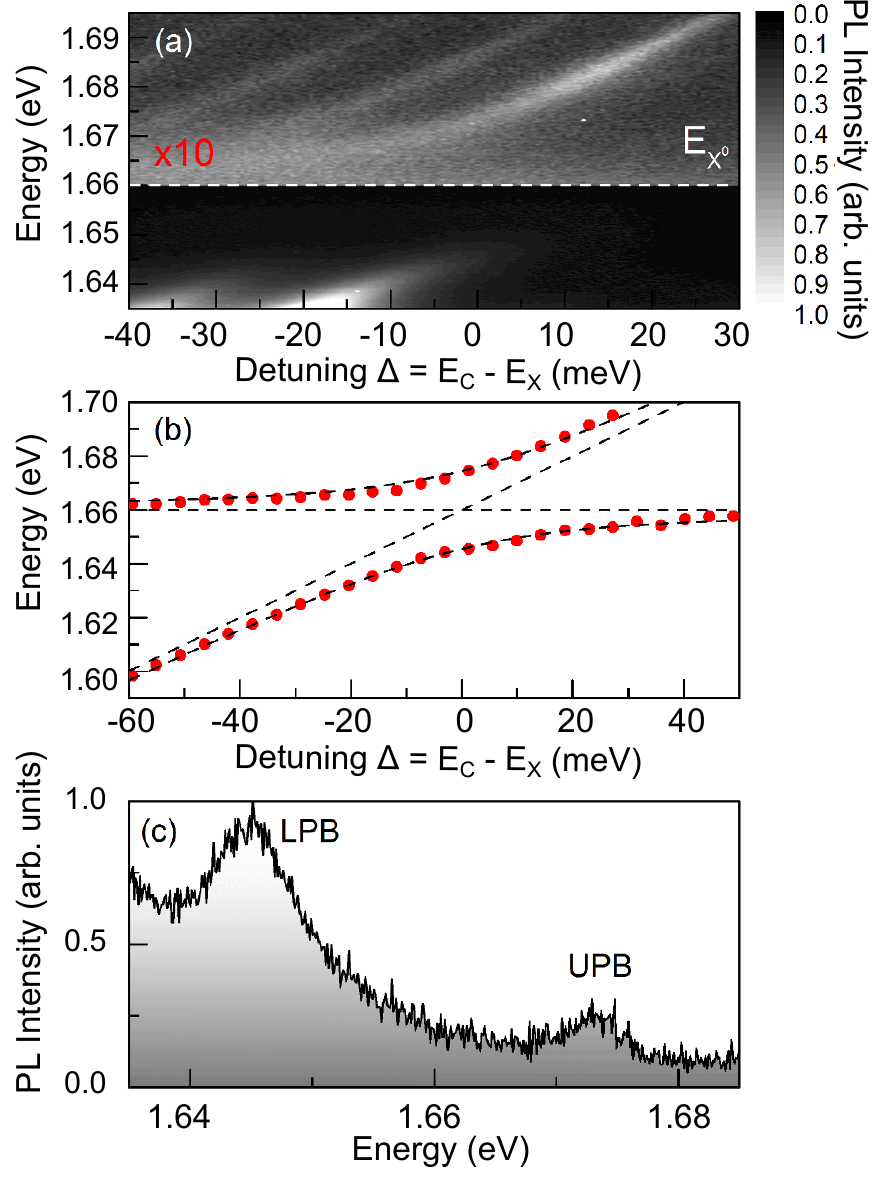}
\caption{\label{fig4} {\bf Observation of polariton states in a double quantum well heterostructure.} (a) The double QW structure shows an anticrossing between the neutral exciton and discrete cavity modes at 4 K. (b) A fit to the peak position as a function of detuning yields a Rabi splitting of 29 meV. (c) The upper and lower polariton branches are well resolved at resonance.}
\end{figure}

Figure~\ref{fig4}(a) shows the PL spectra from a cavity formed with the double QW region of the heterostructure as the active region. Here an anticrossing is observed between the cavity and neutral exciton resonance where a fit to the polariton peak energies using the coupled oscillator model, shown in Figure~\ref{fig4}(b), reveals an increased Rabi splitting of 29 meV. This increase is in agreement with the expected scaling of the Rabi splitting with QW number $N_{QW}$ of $\hbar\Omega_{Rabi} \propto \sqrt{N_{QW}}.$\cite{Kavokin2007} The spectral slice at resonance is shown in Figure~\ref{fig4}(c) where the labeled polariton resonance can be fully resolved. We anticipate many applications of multiple QW heterostructures where, due to the large exciton binding energy, stable room temperature (RT) polaritons can be expected. A heterostructure consisting of 4 MoSe$_2$ QWs is expected to exhibit a Rabi splitting of at least $40$ meV, larger than the room temperature exciton linewidth of 35 meV (see Supplementary Information). Narrow polariton resonances at room temperature will require that half of the Rabi splitting is larger than the width of the exciton resonance. The fabrication of VDW heterostructures with large numbers of QWs should allow the fulfilment of this criterion.

In summary we have conclusively demonstrated strong exciton-photon coupling of MoSe$_2$ heterostructures in tunable optical microcavities through the observation of an anticrossing with the neutral exciton energy in PL and reflectivity. For a single MoSe$_2$ monolayer a Rabi splitting of 20 meV is observed for X$^0$. We also observe an intermediate coupling regime with X$^-$, present due to the inherent doping from impurities in the monolayer films with an estimated coupling strength of 8.2 meV. We extended this to the demonstration of multiple QW TMDC heterostructures in the strong coupling regime where the Rabi splitting is increased to 29 meV. Unique to TMDCs microcavities is the presence of a bright and narrow UPB due to the large exciton binding energy causing the electron-hole continuum to be far from the polariton resonance. This will allow the potential realisation of phenomena involving the UPB, such as highly non-linear parametric processes with balanced idler/signal, or polariton quantum-bits\cite{Demirchyan2014}. Other avenues of van der Waals crystal based polaritonics may include studies involving the spin-valley coupling between excitonic states which leads to the formation of LT-polarised excitons with potential large polarisation splitting that will be inherited by the polaritonic system\cite{Glazov2014}. This type of spin-orbit interaction is at the origin of the optical spin Hall effect \cite{Leyder2007} and the formation of a persistent spin currents in polariton condensates\cite{Sala2015}. It allows manipulation of polariton trajectories through the emergence of non-Abelian gauge fields \cite{Tercas2014}. Lattices of coupled open cavities \cite{Flatten2015} filled with VDW heterostructures could be realized in the near future and serve as a basis for the realization of polaritonic topological insulators \cite{Nalitov2015_2} with room temperature operation.

\section{Methods}

\subsection{Dielectric mirror fabrication}

The concave shaped template for the top mirror is produced by focused ion beam milling, where an array of concave shaped mirrors with radii of curvatures ranging from 7 to 20 \micro m is milled in a smooth fused silica substrate. Gallium ions are accelerated onto a precise position of the silica substrate achieving an accuracy of around 5 nm with an rms roughness below 1 nm.\cite{Dolan2010} The highly reflecting distributed Bragg reflectors are coated by ion assisted electron beam deposition, where 10 pairs of quarter-lambda SiO$_2$/NbO$_2$ layers (refractive index 1.4 and 2.0 respectively), designed for a center wavelength of 750 nm and a stop-band width of around 200 nm, are deposited on the substrates. 

\subsection{2D film fabrication}

The monolayer sheets of MoSe$_2$ and the thin films of hexagonal boron nitride (hBN) were obtained by mechanical exfoliation of bulk crystals. The first monolayer is transferred onto a thin layer of hexagonal boron nitride (hBN) using standard transfer techniques \cite{KretininNanoLett2014}. This step is repeated with the 3 nm thick sheet of hBN and the second monolayer sheet of MoSe$_2$. The final heterostructure, consisting of a single monolayer region, a double QW region and a MoSe$_2$ bilayer region was then transferred onto the planar dielectric mirror. The individual regions can be identified using optical imaging. Bulk crystals were acquired from HQGraphene.

\subsection{Optical measurements}

Optical measurements were performed with the samples placed in a helium bath cryostat system at a temperature of 4.2K. Top and bottom mirrors were attached to attocube closed-loop XYZ nano-positioners and a tilt positioner allowing independent sample positioning with a travel range of 5 mm with a few tens of pm precision. The optical properties of the monolayer MoSe$_2$ can be measured when the top mirror is moved out of the optical path using the lateral translation stages. A plano-concave microcavity is formed when the concave mirror is brought back into the optical path. All \micro-PL experiments were performed with a continuous-wave (cw) excitation using a 638 nm laser diode, focused onto the sample using an achromatic lens. The collected PL is focused onto a wound fiber bundle and guided into a 0.75 m spectrometer and a high sensitivity charge coupled device for emission collection. Time-resolved measurements are obtained using a picosecond pulsed, frequency doubled titanium-sapphire laser with a pulse length of around 3 ps. The exponential decay of the monolayer emission is collected using a streak camera.
\newline

{\bf ACKNOWLEDGMENTS \\}

We thank the financial support of the Graphene Flagship, FP7 ITN S$^3$NANO, ERC grant EXCIPOL 320570, and the EPSRC Programme Grant EP/J007544/1. O.D.P.Z gratefully thanks CONACYT-Mexico. A.A.P.T. and J.M.S. acknowledge support from the Leverhulme Trust. F. W. acknowledges support from the Royal Academy of Engineering and K. S. N. from the Royal Society and ERC grant Hetero2D. \\ 

{\bf AUTHOR CONTRIBUTIONS\\}

S. D. and  S. S. carried out optical investigations with contributions from F. L., M. S. and O. D.-P. Z. TMDC heterostructures were fabricated by F. W and A. A. P. T. designed and fabricated the concave mirrors. C. C. deposited the DBRs. A. N., D. D. S. and G. M. carried out the theoretical analysis. S. D. and S. S. analyzed the data. S. D. and S. S. wrote the manuscript with contributions from all co-authors. K. S. N., J. M. S., M. S. S., D. N. K. and A. I. T. provided management of various aspects of the project, and contributed to the analysis and interpretation of the data and writing of the manuscript. A. I. T. conceived and oversaw the whole project.

\newpage

{\bf MANUSCRIPT FIGURES\\}

\widetext
\clearpage
\begin{center}
\textbf{\large Supplementary Information for ``Spin Textures of Polariton Condensates in a Tunable Microcavity with Strong Spin-Orbit Interaction''}
\end{center}
\renewcommand{\thesection}{S\arabic{section}}
\setcounter{section}{0}
\renewcommand{\thefigure}{S\arabic{figure}}
\setcounter{figure}{0}
\renewcommand{\theequation}{S\arabic{equation}}
\setcounter{equation}{0}
\renewcommand{\thetable}{S\arabic{table}}
\setcounter{table}{0}
\renewcommand{\citenumfont}[1]{S#1}
\makeatletter
\renewcommand{\@biblabel}[1]{S#1.}
\makeatother

\onecolumngrid

\maketitle

In this Supplementary Materials we present additional details on the strong exciton-photon coupling including reflectivity measurements, studies on bilayer molybdenum diselenide, room temperature measurements and studies on the polariton intensity. Theoretical models support our observations and fit well with the measured data.

\section{\label{SI:Theory}Theoretical calculation of the expected Rabi splitting}

Coupling a photonic open cavity mode to an exciton level is characterized by the Rabi frequency \cite{Kavokin2007}:

\begin{equation}
\Omega_{Rabi} = 2\sqrt{\frac{2\Gamma_0c}{n_c(L_{DBR}+L_c)}},
\label{eq1}
\end{equation}

\noindent
where $n_c$ is the cavity refractive index, which is close to unity in the open cavity system, $L_{DBR}$ and $L_c$ are the effective mirror and cavity length length respectively. $\Gamma_0=1/2\tau$ is the exciton radiative broadening given by \cite{Ivchenko2005}:

\begin{equation}
\Gamma_0=\frac{2\pi e^2|p_{cv}|^2}{nc\hbar \omega_0 m_e^2}\phi(\rho)^2,
\label{eq2}
\end{equation}

\noindent
where $e$ is the electron charge absolute value, $p_{cv}$ is the matrix element of the momentum between electron Bloch functions at valence and conduction band edges, $n\approx 2.2$ is the refractive index of MoSe$_2$, $c$ is the speed of light, $\hbar \omega_0$ is the cavity mode energy, $m_e$ is the free electron mass and $\phi(\rho)$ is the internal motion part of the 2D exciton wavefunction. For the 1s exciton state one can write:

\begin{equation}
\phi(0) = \sqrt{\frac{2}{\pi a_B^2}} = \sqrt{\frac{2\mu E_b}{\pi \hbar^2}},
\end{equation}

\noindent
with $a_B$ the 2D Bohr radius, $E_b \approx 0.55$ eV the exciton binding energy and $\mu = m_e \frac{m_c^*m_v^*}{m_c^*+m_v^*}$ the exciton reduced mass.

The matrix element $p_{cv}$ may be deduced from the electron effective mass expression given by the $k\cdot p$ method:

\begin{equation}
m_c^* = \frac{m_c}{m_e} = \left( 1+\frac{2|p_{cv}|^2}{E_gm_e} \right)^{-1},
\end{equation}

\noindent
where $E_g\approx 2.1$ eV is the band gap. Conduction and valence band effective masses are calculated ab initio for MoSe$_2$ and are given by $m_{c(v)}^*=0.70(0.55)$ \cite{Nalitov2015}. From this we derive:

\begin{equation}
|p_{cv}| = \sqrt{\frac{E_gm_e}{2}\left(\frac{1}{m_c^*}-1\right)}.
\label{eq5}
\end{equation}

Substituting equations \ref{eq2} and \ref{eq5} in equation \ref{eq1} and assuming no detuning between exciton energy and photonic mode $\left( \hbar \omega_0 = E_g - E_b \right)$ we obtain the Rabi splitting:

\begin{equation}
\hbar \Omega_{Rabi} = 8\sqrt{\frac{m_v^*\left( 1-m_c^* \right)}{2n\left( m_c^* + m_v^* \right)}\frac{E_gE_b}{E_g-E_b}\frac{e^2}{L_{DBR}+L_c}}.
\end{equation}

With the absolute effective cavity length $L_{DBR} + L_c = 2.3$ $\mu$m , the obtained Rabi splitting is $\hbar \Omega_{Rabi} \approx 26.7$ meV. This agrees well with the experimentally obtained value of 20 meV.

\section{\label{SI:Lifetime}Calculation of the exciton radiative lifetime}

The obtained Rabi splitting for a single monolayer sheet is $\Omega_{Rabi}=20$ meV. Following equation \ref{eq1}, the exciton radiative rate $\Gamma_0$ and therefore the radiative lifetime $\tau$ can be obtained using

\begin{equation}
\Omega_{Rabi} = \frac{20 \text{ meV}}{\hbar} = 2\sqrt{\frac{2\Gamma_0c}{n_c(L_{DBR}+L_c)}}.
\end{equation}

The total cavity length is determined by the free spectral range between two longitudinal modes $(L_{DBR} + L_c) = 2.3$ $\mu$m, $c$ is speed of light and $n_c = 1.4$ is the effective cavity refractive index. This allows the exciton radiative lifetime to be calculated to be $\Gamma_0 = \frac{1}{0.8\text{ ps}}$. With $\Gamma_0 = \frac{1}{2\tau}$ the exciton radiative lifetime is then $\tau = 0.4$ ps. This is around 13x faster than the exciton lifetime of 5.3 ps measured in Figure 1 (d) of the main text which is determined by relaxation effects to low k-states. The homogeneous exciton linewdith is then given by $\Delta E = \hbar/\tau =$ 1.6 meV. This is much smaller than the low temperature PL linewidth of 11 meV indicating that significant broadening occurs due to disorder effects, an aspect which might be improved using epitaxial layers allowing much narrower polariton linewidths.

\section{\label{SI:RabiBinding}Comparison with other material systems}

Comparison of various material systems in which strong coupling has been observed.

\begin{center}
\begin{tabular} {| l | c | c | c | c |}
	\hline
	Material & Binding energy in bulk (QWs) (E$_B$) & Rabi splitting ($\Omega_{Rabi}$) & $\Omega_{Rabi}/E_B$ & Polariton linewidth at resonance \\ \hline
	
	GaAs & 4.8 ( $\sim$ 14) meV \cite{Atanasov1994,Yu2005} & $\sim$ 3-15 meV \cite{Weisbuch1992,Wertz2010} & $\sim$ 0.2-1.1 & $\sim$ 0.1-1 meV \\ \hline
	
	CdTe & 10 ( $\sim$ 25) meV \cite{Andre1998,Yu2005} & $\sim$ 16-26 meV \cite{Andre1998,Kasprzak2006} & $\sim$ 0.6-1 & $\sim$ 0.6 meV \cite{Kasprzak2006} \\ \hline
	
	GaN & 26 ( $\sim$ 40) meV \cite{Kornitzer1999,Christman2008} & $\sim$ 50 meV \cite{Christman2008} & $\sim$ 1 & $\sim$ 10-15 meV \cite{Christman2008_2,Christopoulos2007}\\ \hline
	
	ZnO & 60 meV\cite{Klingshirn2010} & $\sim$ 200 meV \cite{Li2013} & $\sim$ 3 & $\sim$ 1.5-10 meV \cite{Li2013} \\ \hline
	
	Organics & 250 - 500 meV \cite{Lidzey1998} & $\sim$ 110 meV \cite{Lidzey1998} & $\sim$ 0.2-0.5 & $\sim$ 16.5-25 meV \cite{Lidzey1998,KenaCohen2010} \\
		        & 1 eV \cite{KenaCohen2010} & $\sim$ 200 meV \cite{KenaCohen2010} & 					 &  \\ \hline
	
	MoSe$_2$ & 470 meV \cite{Berkelbach2013} & 20-29 meV & $\sim$ 0.04 & $\sim$ 4-9 meV \\ \hline
\end{tabular}
	\label{table:T1}
\end{center}

\section{\label{SI:Bilayer}Studies on bilayer molybdenum diselenide sheets}

\begin{figure*}[htdp]
\includegraphics[scale=0.85]{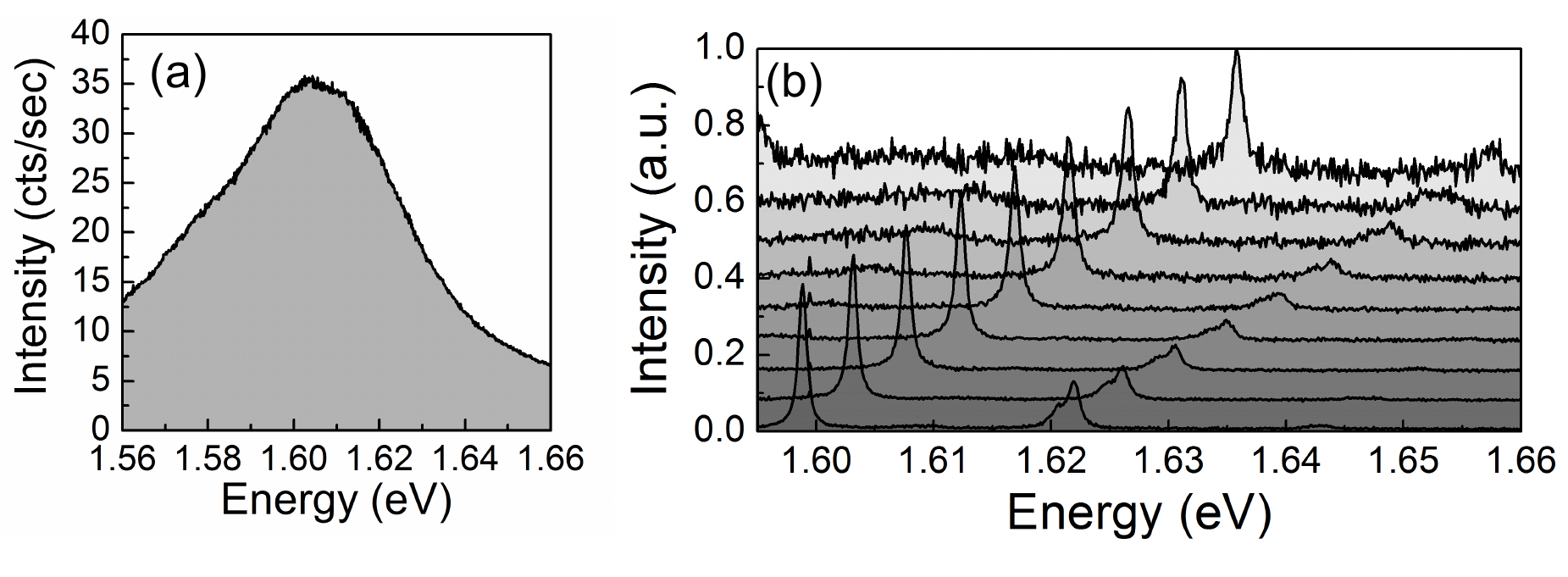}
\caption{\label{figSI3} {\bf Crossing between the cavity modes and bilayer heterostructure area.} (a) Bilayer emission due to indirect transition at 4 K. (b) Spectra of the cavity emission with a bilayer active region showing weak coupling.}
\end{figure*}

The PL emission of the bilayer area marked by the black border in Figure 1 (b) in the main text is shown in Figure~\ref{figSI3}(a). The bandstructure of bilayer MoSe$_2$ shows that there is an indirect transition causing a reduction in the emission efficiency. When the bilayer is coupled to the cavity modes, weak coupling is observed as shown in Figure~\ref{figSI3}(b) where a crossing through the exciton is observed.

\section{\label{SI:Roomtemperature}Room temperature measurements}

The PL spectrum of X$^0$ at room temperature shows a linewidth of around 35 meV, exceeding the vacuum Rabi splitting of both the single and double QW heterostructure. Therefore weak coupling is observed at room temperature. The demonstrated dependence of $\Omega_{rabi} \propto \sqrt{N_{QW}}$ indicates that a heterostructure consisting of four or more MoSe$_2$ QWs will increase the Rabi splitting sufficiently to resolve both the UPB and LPB at room temperature.

\begin{figure*}[htdp]
\includegraphics[scale=1]{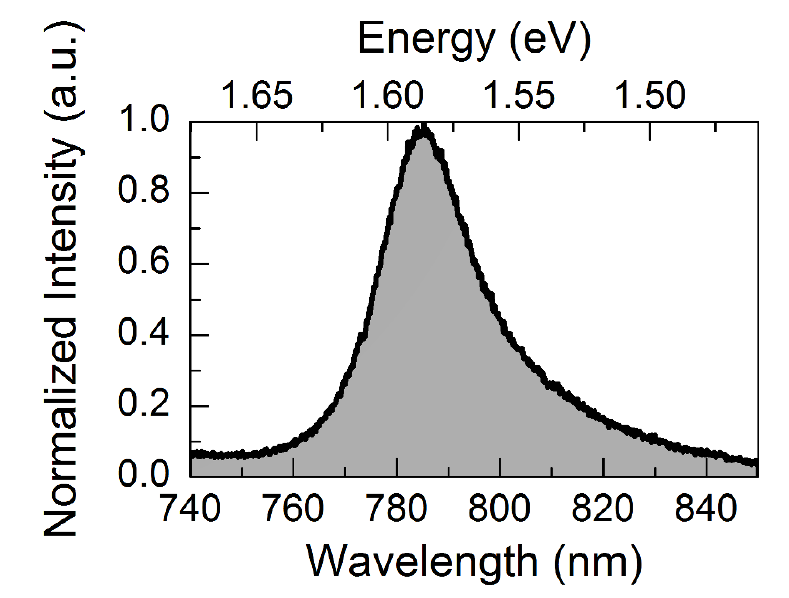}
\caption{\label{figSI4} {\bf Room temperature PL.} A single broad resonance of $X^0$ is observed at room temperature.}
\end{figure*}

\section{\label{SI:Intensities}Polariton intensity}

\begin{figure*}[htdp]
\includegraphics[scale=1.0]{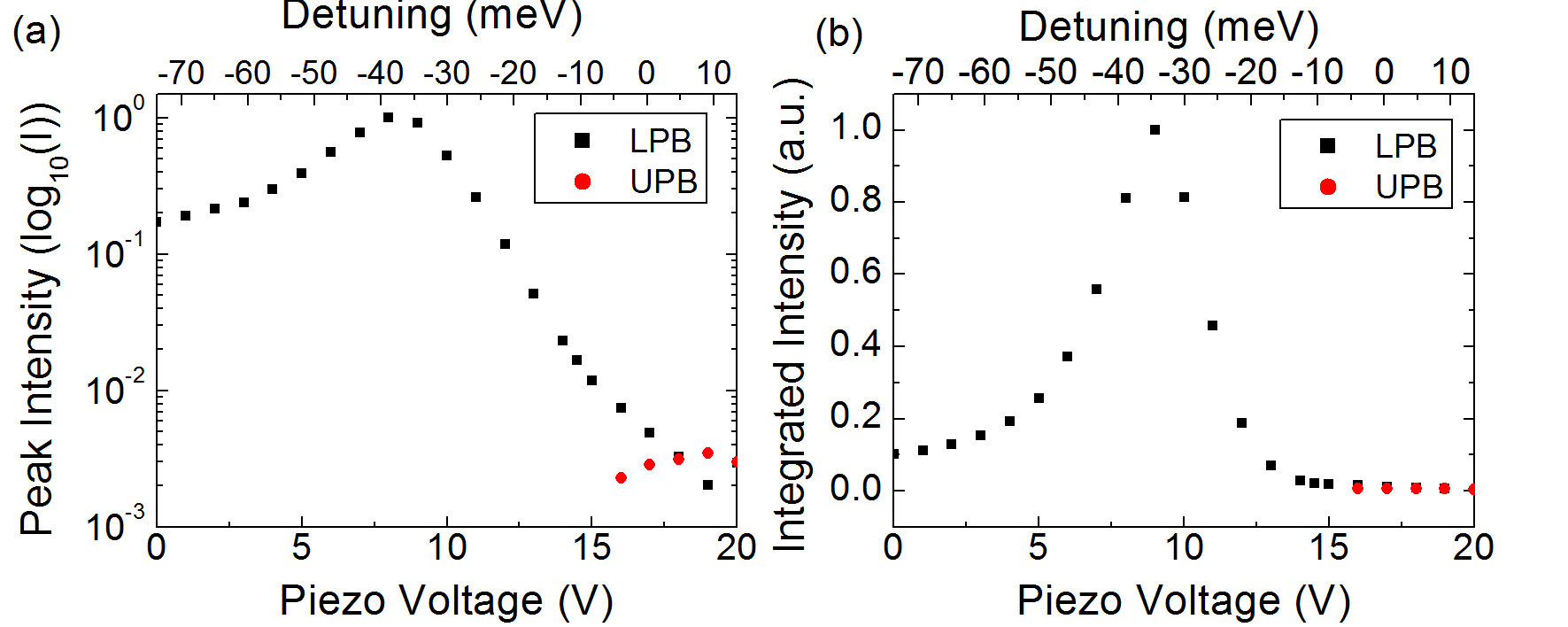}
\caption{\label{figSI5} {\bf Polariton intensity} (a) Peak intensity of the longitudinal resonance polariton as a function of piezo voltage. The detuning is estimated from the coupled oscillator model fit in Figure 2 of the main text. (b) Integrated intensity as a function of piezo voltage.}
\end{figure*}

Figure~\ref{figSI5} (a) shows the peak polariton intensity plotted against longitudinal mode detuning. In resonance with $X^-$ ($\delta = - 30$ meV) the peak intensity is 2 orders of magnitude larger than when in resonance with $X^0$. The integrated intensity is shown in Figure~\ref{figSI5} (b) showing a significant integrated PL enhancement when in resonance with $X^-$. We attribute this to the interplay between the maximum polariton intensity occurring at negative detunings \cite{Stanley1996} as well as intermediate coupling close to $X^-$.  

\begin{thebibliography}{47}
  
\bibitem{Dean2010} {Dean, C. R. {\it et al.} Boron nitride substrates for high-quality graphene electronics, {\it Nature Nanotech.} {\bf 5} 722-726 (2010)}

\bibitem{Britnell2013} {Britnell, L. {\it et al.} Strong Light-Matter Interactions Thin Films, {\it Science} {\bf 340} 1311-1314 (2013)}
  
\bibitem{Geim2013} {Geim, A. K. \& Grigorieva, I. V. Van der Waals heterostructures, {\it Nature} {\bf 499} 419-425 (2013)}

\bibitem{MakPRL2010} {Mak, K., Lee, C., Hone, J., Shan, J., \& Heinz, T. {Atomically Thin MoS2: A New Direct-Gap Semiconductor}, {\it Phys. Rev. Lett.} {\bf 105} 2-5 (2010)}
  
\bibitem{NovoselovPNAS2005} {Novoselov, K. S. {\it et al.} Two-dimensional atomic crystals, {\it PNAS} {\bf 102} 10451-10453 (2005)}
  
\bibitem{WangNatNano2012} {Wang, Q. H., Kalantar-Zadeh, K., Kis, A., Coleman, J. N., \& Strano, M. S. {Electronics and optoelectronics of two-dimensional transition metal dichalcogenides}, {\it Nature Nanotech.} {\bf 7} 699-712 (2012)}  
   
\bibitem{XuNatPhys2014} {Xu, X., Yeo, W., Xiao, D. \& Heinz, T. F. {Spin and pseudospins in layered transition metal dichalcogenides}, {\it Nature Physics} {\bf 10} 343-350 (2014)}  

\bibitem{Ugeda2014} {Ugeda, M. M. {\it et al.} {Giant bandgap renormalization and excitonic effects in a monolayer transition metal dichalcogenide semiconductor}, {\it Nature Mater.} {\bf 13} 1091-1095 (2014)}  

\bibitem{Berkelbach2013} {Berkelbach, T. C., Hybertsen, M. S. \& Reichman, D. R. {Theory of neutral and charged excitons in monolayer transition metal dichalcogenides}, {\it Phys. Rev. B} {\bf 88} 045318 (2013)}  

\bibitem{Mak2012} {Mak, K. F., He, K., Shan, J. \& Heinz, T. F. {Control of valley polarization in monolayer MoS2 by optical helicity}, {\it Nature Nanotech.} {\bf 7} 494-498 (2012)}  
  
\bibitem{Ross:2014fk} {Ross, J. S. {\it et al.} {Electrically tunable excitonic light-emitting diodes based on monolayer WSe2 p-n junctions}, {\it Nature Nanotech.} {\bf 79} 268-272 (2014)}   
  
\bibitem{Rivera2015} {Rivera, P. {\it et al.} {Observation of long-lived interlayer excitons in monolayer MoSe2--WSe2 heterostructures}, {\it Nature Commun.} {\bf 6} 6242 (2015)}     
  
\bibitem{Withers2015} {Withers, F. {\it et al.} {Light-emitting diodes by band-structure engineering in van der Waals heterostructures}, {\it Nature Mater.} {\bf 14} 301-306 (2015)}     
  
\bibitem{Dufferwiel2014} {Dufferwiel, S. {\it et al.} {Strong exciton-photon coupling in open semiconductor microcavities}, {\it App. Phys. Lett.} {\bf 104} 192107 (2014)}   

\bibitem{Schwarz2014} {Schwarz, S. {\it et al.} {Two-Dimensional Metal-Chalcogenide Films in Tunable Optical Microcavities}, {\it Nano Lett.} {\bf 14} 7003-7008 (2014)}   

\bibitem{Gan2013} {Xuetao, G. {\it et al.} {Controlling the spontaneous emission rate of monolayer MoS2 in a photonic crystal nanocavity}, {\it App. Phys. Lett.} {\bf 103} 181119 (2013)}   

\bibitem{Wu2014} {Wu, S. {\it et al.} {Control of two-dimensional excitonic light emission via photonic crystal}, {\it 2D Mater.} {\bf 1} 011001 (2014)}   

\bibitem{Wu2015} {Wu, S. {\it et al.} {Ultra-Low Threshold Monolayer Semiconductor Nanocavity Lasers}, {\it Nature} {\bf 520} 520 (2015)}   

\bibitem{Cheiwchanchamnangij2012} {Cheiwchanchamnangij, T. \& Lambrecht, W. R. L. {Quasiparticle band structure calculation of monolayer, bilayer, and bulk MoS${}_{2}$}, {\it Phys. Rev. B} {\bf 85} 205302 (2012)}   

\bibitem{Jiang2014} {Jian-Hua, J. \& Sajeev, J. {Photonic Architectures for Equilibrium High-Temperature Bose-Einstein Condensation in Dichalcogenide Monolayers}, {\it Sci. Rep.} {\bf 4} 7432 (2014)}   

\bibitem{Liu2015} {Liu, X. {\it et al.} {Strong light-matter coupling in two-dimensional atomic crystals}, {\it Nature Phot.} {\bf 9} 30-34 (2015)}   

\bibitem{Kasprzak2006} {Kaspzak, K. {\it et al.} {Bose-Einstein condensation of exciton polaritons}, {\it Nature} {\bf 443} 409-414 (2006)}   

\bibitem{Amo2009_2} {Amo, A. {\it et al.}  {Superfluidity of polaritons in semiconductor microcavities}, {\it Nature Physics} {\bf 5} 805-810 (2009)}   

\bibitem{Christopoulos2007} {Christopoulos, S. {\it et al.}  {Room-Temperature Polariton Lasing in Semiconductor Microcavities}, {\it Phys. Rev. Lett.} {\bf 98} 126405 (2007)} 

\bibitem{Bhattacharya2014} {Bhattacharya, P. {\it et al.}  {Room Temperature Electrically Injected Polariton Laser}, {\it Phys. Rev. Lett.} {\bf 112} 236802 (2014)} 

\bibitem{Lu2012} {Lu, T-C. {\it et al.} {Room temperature polariton lasing vs. photon lasing in a ZnO-based hybrid microcavity}, {\it Opt. Express} {\bf 20} 5530-5537 (2012)} 

\bibitem{Li2013} {Li, F. {\it et al.} {From Excitonic to Photonic Polariton Condensate in a ZnO-Based Microcavity}, {\it Phys. Rev. Lett.} {\bf 110} 196406 (2013)} 

\bibitem{KenaCohen2010} {Kena-Cohen, S. \& Forrest, S. R. {Room-temperature polariton lasing in an organic single-crystal microcavity}, {\it Phys. Rev. Lett.} {\bf 4} 371-375 (2010)} 

\bibitem{Daskalakis2014} {Daskalakis, K. S. \& Maier, S. A. and Murray, R. and KŽna-Cohen, S. {Nonlinear interactions in an organic polariton condensate}, {\it Nature Mater.} {\bf 13} 271-278 (2014)} 

\bibitem{Plumhof2014} {Plumhof, J. D., Stoferle, T.,  Mai, L.,  Scherf, U. \& Mahrt, R. F. {Room-temperature Bose-Einstein condensation of cavity exciton-polaritons in a polymer}, {\it Nature Mater.} {\bf 13} 247-252 (2014)} 

\bibitem{Dolan2010} {Dolan, P. R., Hughes, G. M., Grazioso, F., Patton, B. R. \& Smith, J. M. {Femtoliter tunable optical cavity arrays}, {\it Nature Communications} {\bf 35} 3556-3558 (2010)} 

\bibitem{RossNatComm2013} {Ross, J. S. {\it{et al.}} {Electrical control of neutral and charged excitons in a monolayer semiconductor}, {\it Optics Lett.} {\bf 35} 1474 (2010)} 

\bibitem{Wang2012} {Wang, Q. H., Kalantar-Zadeh, K., Kis, A., Coleman, J. N. \& Strano, M. S. {Electronics and optoelectronics of two-dimensional transition metal dichalcogenides}, {\it Nature Nanotech.} {\bf 7} 699-712 (2012)} 

\bibitem{Whittaker1998} {Whittaker, D. M. {What Determines Inhomogeneous Linewidths in Semiconductor Microcavities?}, {\it Phys. Rev. Lett.} {\bf 80} 4791-4794 (1998)} 

\bibitem{Kavokin1998} {Kavokin, A. V. {Motional narrowing of inhomogeneously broadened excitons in a semiconductor microcavity: Semiclassical treatment}, {\it Phys. Rev. B} {\bf 57} 3757-3760 (1998)} 

\bibitem{Savona1997} {Savona, V. \& Piermarocchi, C. {Microcavity Polaritons: Homogeneous and Inhomogeneous Broadening in the Strong Coupling Regime}, {\it Physica Status Solidi (a)} {\bf 164} 45-51 (1997)} 

\bibitem{Moody2014} {Moody, G. {\it{et al.}} {Intrinsic Exciton Linewidth in Monolayer Transition Metal Dichalcogenides Intrinsic Exciton Linewidth in Monolayer Transition Metal Dichalcogenides}, {\it arXiv:1410.3143} (2014)} 

\bibitem{Savona1995} {Savona, V., Andreani, L.C., Schwendimann, P. \& Quattropani, A. {Quantum well excitons in semiconductor microcavities: Unified treatment of weak and strong coupling regimes}, {\it Solid State Communications} {\bf 93} 733-739 (1995)} 

\bibitem{Rapaport2001} {Rapaport, R., Cohen, E. Ron, A., Linder, E. \& Pfeiffer, L. N. {Negatively charged polaritons in a semiconductor microcavity}, {\it Phys. Rev. B} {\bf 63} 235310 (2001)} 

\bibitem{Kavokin2007} {Kavokin, A. V., Baumberg, J. J., Malpuech, G. \& Laussy, F. P. {Microcavities}, {\it Oxford University Press} Series on Semiconductor Science and Technology (2007)} 

\bibitem{Demirchyan2014} {Demirchyan, S. S., Chestnov, I. Y.,  Alodjants, A. P., Glazov, M. M., Kavokin, A.V. {Qubits Based on Polariton Rabi Oscillators}, {\it Phys. Rev. Lett.} {\bf 112} 196403 (2014)} 

\bibitem{Glazov2014} {Glazov, M. M. {\it{et al.}} {Exciton fine structure and spin decoherence in monolayers of transition metal dichalcogenides}, {\it Phys. Rev. B} {\bf 89} 201302 (2014)} 

\bibitem{Leyder2007} {Glazov, M. M. {\it{et al.}} {Observation of the optical spin Hall effect}, {\it Nature Physics} {\bf 3} 628-631 (2007)} 

\bibitem{Sala2015} {Sala, V. G. {\it{et al.}} {Spin-Orbit Coupling for Photons and Polaritons in Microstructures}, {\it Phys. Rev. X} {\bf 5} 011034 (2015)} 

\bibitem{Tercas2014} {Tercas, H., Flayac, H., Solnyshkov, D. D. \& Malpuech, G. {Non-Abelian Gauge Fields in Photonic Cavities and Photonic Superfluids}, {\it Phys. Rev. Lett.} {\bf 112} 066402 (2014)} 

\bibitem{Flatten2015} {Flatten, L. C., Trichet, A. A. P., \& Smith, J. M. {Spectral engineering of coupled open-access microcavities}, {\it arXiv:1503.07687} (2014)} 

\bibitem{Nalitov2015_2} {Nalitov, A. V., Solnyshkov, D. D. \& Malpuech, G. {Polariton Z Topological Insulator}, {\it Phys. Rev. Lett.} {\bf 114} 116401 (2015)} 

\bibitem{KretininNanoLett2014} {Kretinin, A. V. {\it{et al.}} {Electronic Properties of Graphene Encapsulated with Different Two-Dimensional Atomic Crystals}, {\it Nano Lett.} {\bf 14} 3270-3276 (2014)} 
  
\end{thebibliography}

\begin{thebibliography}{19}%

\bibitem{Kavokin2007} {Kavokin, A. V., Baumberg, J. J., Malpeuch, G. \& Laussy, F. P.  {\it Microcavities}, Series on Semiconductor Science and Technology (Oxford University Press, 2007)}

\bibitem{Ivchenko2005} {Ivchenko, E. L. {\it Optical spectroscopy of semiconductor nanostructures} (Alpha Science International Ltd, 2005)} 

\bibitem{Nalitov2015} {Nalitov, A. V. Malpuech, G., Tercas, H. \& Solnyshkov, D. D. {Spin-Orbit Coupling and the Optical Spin Hall Effect in Photonic Graphene}, {\it Phys. Rev. Lett.} {\bf 114} 026803 (2015)} 

\bibitem{Atanasov1994} {Atanasov, R., Bassani, F., D'Andrea, A., \& Tomassini, N. {Exciton properties and optical response in ${\mathrm{In}}_{\mathit{x}}$${\mathrm{Ga}}_{1\mathrm{-}\mathit{x}}$As/GaAs strained quantum wells}, {\it Phys. Rev. B.} {\bf 50} 14381 (1994)} 

\bibitem{Yu2005} {Yu, P. Y., \& Cardona, M. {\it Fundamentals of Semiconductors: Physics and Materials Properties} (Springer, 2005)} 

\bibitem{Weisbuch1992} {Weisbuch, C., Nishioka, M., Ishikawa, A., \& Arakawa, Y. {Observation of the coupled exciton-photon mode splitting in a semiconductor quantum microcavity}, {\it Phys. Rev. Lett.} {\bf 69} 3314 (1992)} 

\bibitem{Wertz2010} {Wertz, E. {\it{et al.}} {Spontaneous formation and optical manipulation of extended polariton condensates}, {\it Nat. Phys.} {\bf 6} 860 (2010)}

\bibitem{Andre1998} {Andre, R., Heger, D., Dang, L. S., \& d'Aubigne, Y. M. {Spectroscopy of polaritons in CdTe-based microcavities}, {\it Journal of Crystal Growth} {\bf 184-185} 758-762 (1998)}

\bibitem{Kasprzak2006} {Kasprzak, E. {\it{et al.}} {Bose-Einstein condensation of exciton polaritons}, {\it Nature} {\bf 443} 409 (2006)}

\bibitem{Kornitzer1999} {Kornitzer, K. {\it{et al.}} {Photoluminescence and reflectance spectroscopy of excitonic transitions in high-quality homoepitaxial GaN films}, {\it Phys. Rev. B} {\bf 60} 1471 (1999)}

\bibitem{Christman2008} {Christmann, G. {\it{et al.}} {Large vacuum Rabi splitting in a multiple quantum well GaN-based microcavity in the strong-coupling regime}, {\it Phys. Rev. B} {\bf 77} 085310 (2008)}

\bibitem{Christman2008_2} {Christmann, G. {\it{et al.}} {Room temperature polariton lasing in a GaN?AlGaN multiple quantum well microcavity}, {\it App. Phys. Lett.} {\bf 93} 051102 (2008)}

\bibitem{Christopoulos2007} {Christopoulos, S. {\it{et al.}} {Room-Temperature Polariton Lasing in Semiconductor Microcavities}, {\it App. Phys. Lett.} {\bf 98} 126405 (2007)}

\bibitem{Klingshirn2010} {Klingshirn, C. F.,  Waag, A., Hoffman, A., \& Geurts, J.  {\it Zinc oxide: from fundamental properties towards novel applications} (Springer 2010)}

\bibitem{Li2013} {Li, F. {\it et al.} {From Excitonic to Photonic Polariton Condensate in a ZnO-Based Microcavity}, {\it Phys. Rev. Lett.} {\bf 110} 196406 (2013)} 

\bibitem{Lidzey1998} {Lidzey, D. G. {\it et al.} {Strong exciton-photon coupling in an organic semiconductor microcavity}, {\it Nature} {\bf 395} 53--55 (1998)} 

\bibitem{KenaCohen2010} {Kena-Cohen, S., \& Forrest, S. R. {Room-temperature polariton lasing in an organic single-crystal microcavity}, {\it Nat. Phot.} {\bf 4} 371--375 (2010)} 

\bibitem{Berkelbach2013} {Berkelbach, T. C., Hybertsen, M. S., \& Reichman, D. R. {Theory of neutral and charged excitons in monolayer transition metal dichalcogenides}, {\it Phys. Rev. B} {\bf 88} 045318 (2013)} 

\bibitem{Stanley1996} {Stanley, R. P. {\it {et al}}. {Cavity-polariton photoluminescence in semiconductor microcavities: Experimental evidence}, {\it Phys. Rev. B} {\bf 53} 10995--11007 (1996)} 

\end{thebibliography}
\end{document}